\documentclass{emulateapj}

\slugcomment{Accepted by the Astrophysical Journal}

\newcommand{\hhh}{H$_3^+$}

\def\msol{M_\odot}
\def\pfrac#1#2{\left(\frac{#1}{#2}\right)}
\def\beq{\begin{equation}}
\def\eeq{\end{equation}}

\def\iso#1#2{\mbox{${}^{#2}{\rm #1}$}}
\def\li#1{\iso{Li}{#1}}
\def\be#1{\iso{Be}{#1}}
\def\bor1#1{\iso{B}{1#1}}

\def\ry{{\rm Ry}}

\begin{document}

\title{The Implications of a High Cosmic-Ray Ionization Rate in Diffuse
Interstellar Clouds}

\author{Nick Indriolo\altaffilmark{1},
Brian D. Fields\altaffilmark{1,2},
Benjamin J. McCall\altaffilmark{1,2,3}}

\altaffiltext{1}{Department of Astronomy, University of Illinois at
Urbana-Champaign, Urbana, IL 61801}
\altaffiltext{2}{Department of Physics, University of Illinois at
Urbana-Champaign, Urbana, IL 61801}
\altaffiltext{3}{Department of Chemistry, University of Illinois at
Urbana-Champaign, Urbana, IL 61801}

\begin{abstract}

Diffuse interstellar clouds show large abundances of \hhh\ which can be maintained only by a high ionization rate of H$_2$.  Cosmic rays are the dominant ionization mechanism in this environment, so the large ionization rate implies a high cosmic-ray flux, and a large amount of energy residing in cosmic rays.  In this paper we find that the standard propagated cosmic-ray spectrum predicts an ionization rate much lower than that inferred from \hhh.
Low-energy ($\sim 10$ MeV) cosmic rays are the most efficient at ionizing hydrogen, but cannot be directly detected; consequently, an otherwise unobservable enhancement of the low-energy cosmic-ray flux offers a plausible explanation for the \hhh\ results. Beyond ionization, cosmic rays also
interact with the interstellar medium by spalling atomic nuclei and exciting atomic nuclear states.  These processes produce the light elements Li, Be, and B, as well as gamma-ray lines.  To test the consequences of an enhanced low-energy cosmic-ray flux, we adopt two physically-motivated cosmic-ray spectra which by construction reproduce the ionization rate inferred in diffuse clouds, and investigate the implications of these spectra on dense cloud ionization rates, light element abundances, gamma-ray fluxes, and energetics.  One spectrum proposed here provides an explanation for the high ionization rate seen in diffuse clouds while still appearing to be broadly consistent with other observables, but the shape of this spectrum suggests that supernovae remnants may not be the predominant accelerators of low-energy cosmic rays.


\end{abstract}


\section{INTRODUCTION}

Several recent observations of \hhh\ in the diffuse interstellar medium (ISM) indicate an average cosmic-ray ionization rate of molecular hydrogen
of about $4\times 10^{-16}$~s$^{-1}$ \citep{mcc03,ind07}.
This value is about 1 order of magnitude larger
than was previously inferred using other molecular tracers such as HD and OH
\citep{odo74,bla77,bla78,har78a,har78b,fed96}.  However, several models have
also required ionization rates on the order of $10^{-16}$ s$^{-1}$
\citep{van86,lis03,lep04,sha06,sha08} in order to reproduce the observed abundances of various atomic and molecular species.  This agreement, coupled with the simplicity behind the chemistry of \hhh, leads us to conclude that the newer measurements are most likely correct.
In this paper, we explore the implications that a high ionization rate has for cosmic rays and related observables.

Aside from observational inferences, the cosmic-ray ionization rate can also
be calculated theoretically using an ionization cross section
and cosmic-ray energy spectrum.  While the ionization cross section for
hydrogen is well determined \citep{bet33,ino71},
the cosmic-ray spectrum below about 1~GeV is unknown.  This is because low energy cosmic rays are deflected from the inner solar system by the magnetic field coupled to the solar wind (an effect called modulation) and so the flux at these energies cannot be directly observed.  This theoretical calculation of the ionization rate has been performed several times \citep[e.g.,][]{hay61,spi68,nat94,web98}, with each study choosing a different low energy cosmic-ray spectrum for various reasons.  Most recently, \citet{web98} predicted an ionization rate of
$(3-4)\times 10^{-17}$~s$^{-1}$ using interstellar proton, heavy nuclei, and electron cosmic-ray spectra.
The proton and heavy nuclei spectra were found by attempting to remove the effects of solar modulation from {\it Pioneer} and {\it Voyager} observations, while the electron spectrum was derived from radio and low-energy gamma-ray measurements.
Even having accounted for all of these components, this result falls about 1 order of magnitude short of the inference based on \hhh, suggesting that the de-modulated solar system spectrum may not be the same as the interstellar spectrum in diffuse clouds, and/or that the \citet{web98} extrapolation to low energies underestimates the true interstellar value.

Together, all of the above studies have shown that the ionization of interstellar hydrogen is a powerful observable for probing cosmic-ray interactions with the environments through which they propagate.
Beyond ionization though, cosmic rays will interact with the ISM in other ways which lead to additional and complementary observables.  Namely, inelastic collisions between cosmic-rays and interstellar nuclei inevitably: (i) create light element isotopes $^6$Li, $^7$Li, $^9$Be, $^{10}$B, and $^{11}$B  when cosmic rays spall C, N, and O nuclei \citep{rev70,men71}, and
(ii) excite nuclear states such as $\iso{C}{12}^*$ and
$\iso{O}{16}^*$, the decay of which
produce gamma-ray lines, most prominently at 4.44~MeV and 6.13~MeV, respectively \citep{men75a}.
Similar to the theoretical calculation of the ionization rate, a
cosmic-ray spectrum and relevant cross sections can be used to determine the
production rates of light elements and gamma-ray lines.  Both the light element
\citep[e.g.,][]{men71,men75b,wal85,ste92,pra93,van96,val02,kne03} and gamma-ray
\citep[e.g.,][]{men75a,ram79,cas95,fie96,tat04} calculations have been performed multiple times, again with each study choosing a different low energy cosmic-ray spectrum.

\begin{figure}
\epsscale{1.25}
\plotone{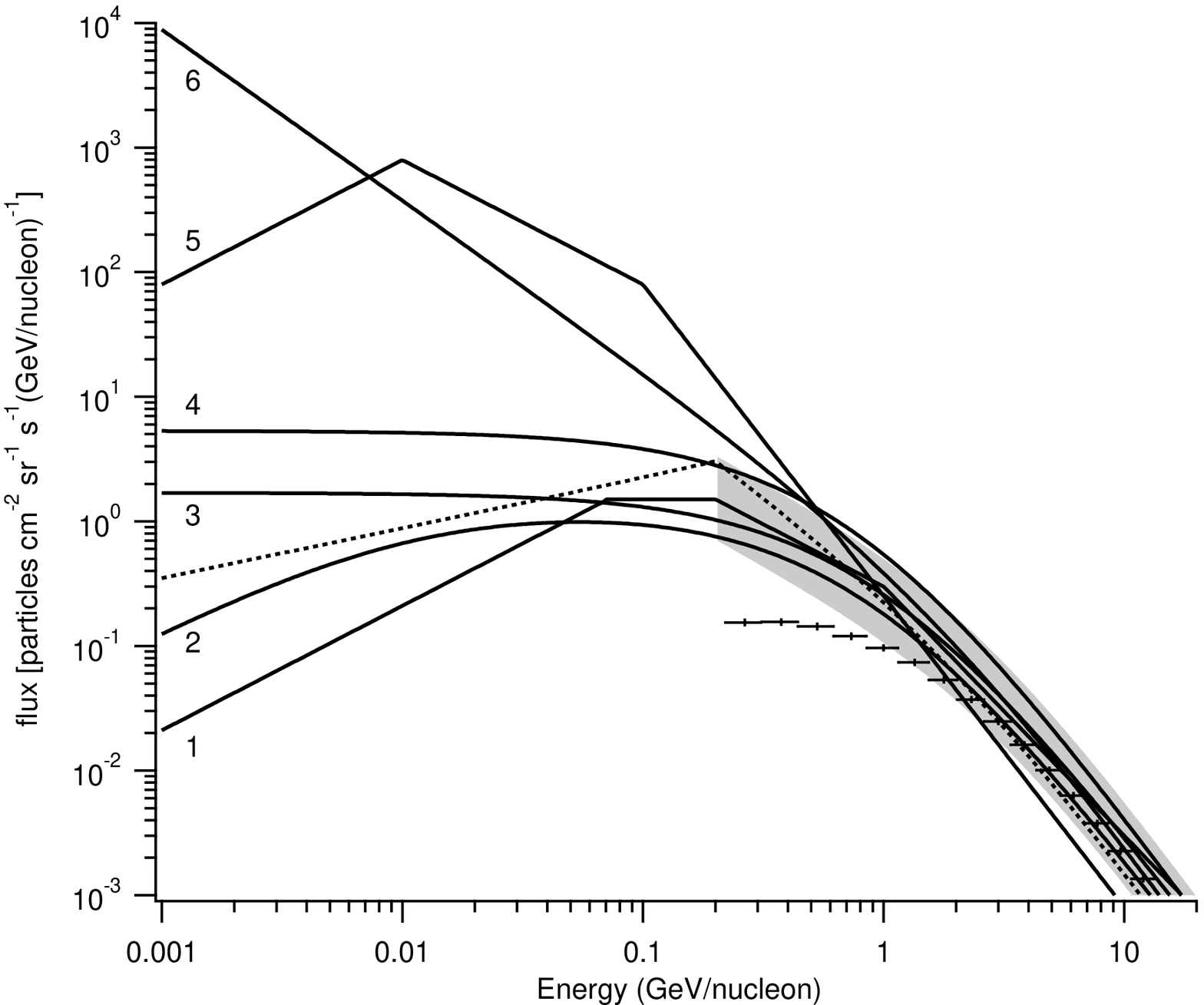}
\caption{Example cosmic-ray spectra used in the literature over the past few
decades.  1 - \citet{her06}; 2 - \citet{spi68}; 3 - \citet{glo69} (via fitting function of \citet{kne03});
4 - \citet{ip85} (via fitting function of \citet{val02}); 5 - \citet{hay61}; 6 - \citet{nat94}.  The dotted line is the propagated leaky box spectrum used in this paper, also shown in Figure \ref{fig2}.  Note the agreement above and discrepancy below 1 GeV.  These spectra were selected to be illustrative of choices in the literature used for different applications.  Of these, it is worth noting that the \citet{ip85} spectrum is the result of a calculation specifically designed to recover the (demodulated) propagated interstellar cosmic-ray spectrum.  The shaded region shows the range of uncertainty in the demodulated proton spectrum as described by \citet{mor97}.  Crosses are proton data from
the top of the Earth's atmosphere
\citep{ams02}
and clearly show the effects of modulation below $\sim 1$ GeV.
}
\label{fig1}
\end{figure}

Some of the spectra that have been used for these calculations are shown in Figure \ref{fig1}.  While several more cosmic-ray spectra have been used, many share functional forms with those plotted and so have been omitted for the sake of clarity.  Note that all of the spectra agree with demodulated data (shaded region) above a few hundred MeV and raw data (crosses) above a few GeV, but that they can differ by about 4 orders of magnitude at 1 MeV.
Figure \ref{fig1} is shown primarily to illustrate our poor understanding of the low-energy portion of the cosmic-ray spectrum.

In this study, we calculate the cosmic-ray ionization rate using several
low energy cosmic-ray spectra in an attempt to reproduce
the value inferred from \hhh\ observations.  For the spectra that
successfully predict ionization rates close to the inferred
value of $4\times10^{-16}$ s$^{-1}$, we further investigate
the implications that they have on dense cloud ionization rates, light element abundances, gamma-ray fluxes, and energetics arguments.

\section{THE IONIZATION RATE INFERRED FROM \hhh}

The chemistry behind \hhh\ in the diffuse ISM is rather simple.  Its formation
and destruction are given by the reactions:
\begin{equation}
{\rm CR} + {\rm H}_2 \rightarrow {\rm CR} + {\rm H}_2^+ + {\rm e}^-,
\label{eqh3+1}
\end{equation}
\begin{equation}
{\rm H}_2 + {\rm H}_2^+ \rightarrow {\rm H}_3^+ + {\rm H},
\label{eqh3+2}
\end{equation}
\begin{equation}
{\rm H}_3^+ + {\rm e}^- \rightarrow {\rm H}_2 + {\rm H}\ {\rm or}\ {\rm H+H+H}.
\label{eqh3+3}
\end{equation}
H$_2$ is first ionized, after which the H$_2^+$ ion
quickly reacts with another H$_2$ molecule to form \hhh.  In diffuse (and dense) clouds it is assumed that this ionization is due almost entirely to cosmic rays, as the flux of photons with $E>13.6$~eV will be quickly attenuated by atomic hydrogen in the outer regions of the cloud.  The first step is many orders of magnitude slower, so it acts as the overall rate limiting step.  The primary channel for destroying \hhh\ in diffuse clouds is recombination with an electron.  \hhh\ is destroyed by reaction (\ref{eqh3+3}) on a time scale of about 100 years, much shorter than the $\sim10^6$~yr lifetime of diffuse clouds \citep{wag88}, so the steady-state approximation is valid and the formation and destruction rates can be equated.
This assumption yields \citep{geb99}
\begin{equation}
\zeta_{2}n({\rm H_2})=k_{e}n({\rm H_3^+ })n(e),
\label{eqh3+ss}
\end{equation}
where $\zeta_2$ is the ionization rate of H$_2$, $k_e$ is the \hhh-electron
recombination rate constant, and $n({\rm X})$'s are number densities.  Spectroscopic observations of transitions from the two lowest rotational levels of \hhh, the only levels populated at the low temperatures of diffuse interstellar clouds, provide the \hhh\ column density.  The cloud path length is then found using the observed hydrogen column density and inferred hydrogen number density.  Dividing the \hhh\ column by the path length gives $n({\rm H_3^+})$, and leaves three variables in the steady state equation:
$k_e$, $\zeta_2$, and $n(e)/n({\rm H_2})$.  However, previous work has shown that the \hhh-electron recombination rate
\citep{mcc03,mcc04} and the electron-to-hydrogen ratio \citep{car96} are
relatively well constrained, leaving $\zeta_2$ as the only free parameter.
Starting from eq.~(\ref{eqh3+ss}) and using various other relationships and assumptions, \citet{ind07} derived an equation for the cosmic-ray ionization rate that depends on observables.
This equation was then used to infer the ionization rate toward several diffuse cloud sight lines.  From all of the sight lines with \hhh\ detections, the average cosmic-ray ionization rate of molecular hydrogen was found to be about $4\times 10^{-16}$~s$^{-1}$
with a maximum uncertainty of about a factor of three either
way (see \S 4.2 of \citet{ind07} for a discussion of the calculations and
uncertainties).

\section{IONIZATION ENERGETICS:  A MODEL-INDEPENDENT LOWER BOUND}

Assuming that the ionization rate above is uniform throughout the diffuse Galactic ISM, it is relatively simple to estimate the total, Galaxy-wide amount of power necessary to produce such a high value.  While this assumption of uniformity is not strictly valid (there are diffuse
clouds with $\zeta_2\lesssim10^{-16}$~s$^{-1}$ and clouds in the Galactic center with $\zeta_2\gtrsim10^{-15}$~s$^{-1}$ \citep{oka05,yus07,got08}), the small sample size of sight lines does not allow for the determination of a meaningful relationship between position and ionization rate.
Despite these fluctuations, if all atomic hydrogen experiences the same ionization rate {\it on average}, then the Galactic luminosity in ionizing cosmic rays, $L_{\rm CR,ionize}$, is given by
\begin{equation}
L_{\rm CR,ionize}=\zeta_{\rm H} \, \overline{\Delta E} \,
  \pfrac{M_{\rm H,diffuse}}{m_{\rm H}} ,
\label{eqCRlum}
\end{equation}
where $\overline{\Delta E}$ is the average energy lost by cosmic rays per
ionization event.  The number of hydrogen atoms in diffuse clouds is the ratio of the mass of all atomic hydrogen in diffuse clouds in the
Galaxy, $M_{\rm H}$, to the mass of a hydrogen atom, $m_{\rm H}$.
The ionization rate of {\em atomic} hydrogen, $\zeta_{\rm H}$, is related
to the ionization rate of {\em molecular} hydrogen (the observable probed
by \hhh) by $2.3\zeta_{\rm H}=1.5\zeta_2$
\citep{gla74}.  The coefficients here are further explained in \S5.

Given the ionization rates from the previous section,
we may place a {\em model-independent lower limit}
on the ionizing cosmic-ray luminosity as follows.
Each ionization event requires a cosmic-ray energy input $\overline{\Delta E} > 13.6$ eV, the ionization potential of atomic hydrogen.  On average cosmic rays will lose more than 13.6 eV in the ionization process though,
so by setting $\overline{\Delta E} = 13.6$ eV in eq.~(\ref{eqCRlum}) we calculate a hard lower limit on the power requirement. The gas mass relevant to  eq.~(\ref{eqCRlum}) is that of all neutral hydrogen in Galactic diffuse clouds, which we take to be $M_{\rm H, diffuse} = 5\times 10^{9} \msol$
(the average of \citet{hen82}, \citet{sod94}, and \citet{mis06}).
This value results in a lower limit to the cosmic-ray luminosity of
\beq
\label{eq:Eion}
L_{\rm CR,ionize} >
  0.11\times10^{51} \ {\rm  erg \ (100~yr)^{-1}} \
  \pfrac{M_{\rm H,diffuse}}{ 5\times 10^{9} \msol}.
\eeq
Note that this cosmic-ray ``energy demand'' is in addition to
the requirements found based on cosmic-ray energy lost as the particles escape the Galaxy.  \citet{fie01} estimated the sum of both contributions,
and found  $L_{\rm CR,tot} \simeq  0.5 \times 10^{51} \ {\rm  erg \ (100~yr)^{-1}}$ which is consistent with eq.~(\ref{eq:Eion})
but also implies that ionization represents a significant part of the cosmic-ray energy budget.

However, eq.~(\ref{eq:Eion}) is only the lower limit to the amount of cosmic-ray energy that goes into ionization.  We can get an actual estimate on the luminosity of ionizing cosmic rays by accounting for molecular hydrogen and by using a more precise value of $\overline{\Delta E}$.  According to \citet{cra78} the average energy lost during an ionization event is about 30 eV, which by itself increases $L_{\rm CR,ionize}$ to
$0.24\times10^{51}$~erg~(100~yr)$^{-1}$.  The inclusion of H$_2$ is more complicated.  The mass of H$_2$ is about the same as that of H in the Galaxy, but most H$_2$ resides in dense molecular clouds \citep{bri90} which do not experience the same cosmic-ray ionization rate as the diffuse ISM \citep{dal06}.  Assuming half of all Galactic H$_2$ experiences the ionization rate used above,
$L_{\rm CR,ionize} \approx 0.33\times10^{51} \ {\rm erg \ (100 \ yr)^{-1}}$, a large fraction of the result found by \citet{fie01}.

As it is currently believed that Galactic cosmic rays are accelerated in supernova remnants (SNR), these results have implications for the efficiency with which supernova mechanical energy is transferred to particle acceleration.
If a typical supernova releases $10^{51}$~erg of mechanical energy
\citep[e.g.,][]{arn87,woo88} and $3\pm 2$ supernovae (SNe) occur each century in the Galaxy \citep{van91,dra99}, then at least 4\% of the energy released in SNe must accelerate the cosmic rays which ionize hydrogen in the ISM.
This efficiency climbs to 12\% if we take the more realistic estimate instead of the lower limit.
However, uncertainties in the supernova rate, supernova energy, and
mass of hydrogen in the Galaxy lead to a total uncertainty of about a factor of
5 either way for this value.  It is important to note though that this
calculation depends only on the cosmic-ray ionization rate, and not on an
adopted form of the cosmic-ray spectrum.  In contrast, calculating the
ionization rate is highly dependent on the cosmic-ray spectrum,
to which we now turn.

\section{POSSIBLE SPECTRA OF LOW-ENERGY COSMIC-RAYS}

Given the well-understood physics of the passage of energetic particles through matter,
the ionization rate completely reflects the spectrum of cosmic rays.
In particular, the ionization cross section (below, eq.~\ref{eqHxsec})
grows towards low energies
as $\sigma_{\rm ion} \sim v^{-2} \sim E^{-1}$
which means that the lowest-energy particles have the strongest
effect on ionization.
Given our lack of direct observational constraints on
cosmic rays at low energies, we will examine the ionization
arising from various possible low-energy behaviors
which are physically motivated and/or
have been suggested in the literature.
Here we summarize in a somewhat pedagogical way
some of the main features of the current understanding of
cosmic-ray acceleration and propagation.

The cosmic-ray spectrum with the strongest physical motivation
(in our view) takes supernova explosions to be the engines of Galactic
cosmic-ray acceleration.
That is, supernovae remnants provide the sites for diffusive shock
acceleration and thus act as cosmic-ray sources.
At these sources, diffusive shock acceleration creates
particles with
spectra which are close to simple
power-laws in (relativistic) momentum $p$.
Specifically, consider the ``test-particle'' limit when particle acceleration has a negligible effect on the shock energy and structure.
In this limit, the cosmic-ray production rate, $q$, per unit volume and
time and per unit interval in relativistic momentum has famously been analytically
shown to be
\citep[e.g.,][]{kry77,bel78,bland78}
\beq
\label{eq:shockaccel}
q_{\rm shock \, accel} = \frac{d{\cal N}_{\rm accel}}{dV \, dt \, dp}
  \ \propto \ p^{-\chi},
\eeq
i.e., a power-law in momentum.
Here the momentum index in the case of strong shocks is $\chi = 2 + 4/{\cal M}^2$,
where the shock Mach number is
${\cal M} = v_{\rm shock}/c_{\rm sound,ism} \gg 1$.
The upshot is that for strong shocks (large ${\cal M}$) as one would find in supernova remnants,
the acceleration power law is just slightly steeper than the flattest (i.e.,
largest at high-energy) limiting power-law spectrum allowed by energy conservation:
$q_{\rm lim} \propto p^{-2}$.
Going beyond the test-particle limit requires nonlinear treatment of the feedback of cosmic
ray energy and pressure on the shock structure and evolution; the study of this
nonlinear shock acceleration remains a vital field, but several groups
\citep[e.g.,][]{kan95,ber99,bla02} find
that the accelerated particles have a spectrum which is roughly similar to
the test-particle result, but which shows some concavity in momentum space,
i.e., the effective spectral index $\chi = d\ln q/d\ln p$
does show a modulation around the constant test-particle value.
Intriguingly, there seems to be agreement on the qualitative result that
the low-energy flux will be higher
than for the test-particle predictions.
Unfortunately, the quantitative results remain at present rather model-dependent.
For the purposes of our analysis, we will simply adopt test-particle
power-law acceleration spectra
as in eq.~(\ref{eq:shockaccel}).
Our results can then be viewed as testing the validity of the
test-particle approximation at low energies.

Once produced at acceleration sites, cosmic rays propagate away, and eventually
are removed either by escape from the Galaxy or by stopping in the ISM due to energy losses (predominantly by energy transfer to the ISM, either ionization or excitation of atoms or molecules). Propagation alters the spectra of cosmic rays from those at the sources. Theoretical treatments of these effects typically make the simplifying assumption of a steady state balance between production and losses.  The resulting ``propagated'' spectrum should represent the flux as seen by an average region of the interstellar medium, far from cosmic-ray sources
\citep[elegantly reviewed in][]{str07}.

A full calculation of cosmic-ray propagation at minimum involves
the particle ``flows'' in energy space; the simplest such treatment is
the classic ``leaky-box'' model which treats the Galaxy as a
medium with sources distributed  homogeneously.  More sophisticated
models account for the inhomogeneous Galaxy and effects such as
diffusion and re-acceleration.
In general, when models include the low-energy regime
\citep[e.g.,][]{ler82,shi06}, they
find that when initially accelerated or ``injected'' spectra are power-laws in momentum,
the resulting propagated spectra are very nearly also power laws, with
fairly abrupt changes of spectral indices at characteristic energy scales (``breaks'') at which one loss mechanism comes to dominate over another.
To fix notation, for our purposes cosmic rays are most usefully
characterized by the propagated cosmic-ray flux (strictly speaking, specific intensity) $\phi(E) = d{\cal N}_{\rm cr}/dA \, dt \, d\Omega \, dE$
per unit energy interval.  For all but the most ultra-high energies, cosmic rays are observed to be isotropic, in which case the flux is related to the cosmic-ray number density $n$ via $4\pi \phi(E) = v \ dn/dE$.  Here $E$ is the cosmic-ray kinetic energy; the total relativistic energy is thus $E_{\rm tot} = E+mc^2$.  Relativistic energy and momentum are related by $E_{\rm tot}^2 = (cp)^2 + (m_p c^2)^2$, and $v/c = cp/E_{\rm tot}$.  Using these, it follows that the flux per unit energy is equal to the particle number density per unit momentum: $4\pi \phi(E) = dn/dp$.  Hence, a number spectrum $dn/dp$ that is a power law in $p$ gives a flux with the same power-law of $p(E)$.

As a result, we characterize possible propagated proton spectra with
a piecewise power law in relativistic momentum $p(E)$:
\begin{equation}
\phi_p(E)\!\! =\!\! \left\{ \begin{array}{ll}
\phi_p(E_1)\left(\frac{p(E)}{p(E_1)}\right)^{\gamma_{\rm high}}\!\!\!\!\!\!,
\; & {\rm if}\, E > E_2 \\
\\
\phi_p(E_1)
\left(\frac{p(E_2)}{p(E_1)}\right)^{\gamma_{\rm high}}\left(\frac{p(E)}{p(E_2)}\right)^{\gamma_{\rm low}}\!\!\!\!\!\!,
\; & {\rm if}\, E_{\rm cut} \leq E \leq E_2 \\
\\
0,
\; & {\rm if}\, E < E_{\rm cut} \\
\end{array}\right.\!\!\!.
\label{eqCRflux}
\end{equation}
Here $E_1=1~{\rm GeV}$ is the arbitrary
energy at which the flux is normalized to fit observations;
following  \citet{mor97} we take
$\phi_p(E_1)= 0.22$ cm$^{-2}$ s$^{-1}$ sr$^{-1}$ GeV$^{-1}$.
This and the observed (high-energy) spectral index $\gamma_{\rm high} \approx -2.7$ fixes the high-energy region of the spectrum.

The low-energy portion of the spectrum is crucial for this paper, and we take
the high/low energy break to be
$E_2=0.2~{\rm GeV}$, which is roughly where ionization losses begin to dominate diffusion and/or escape losses.  The power law index for energies below $E_2$ is $\gamma_{\rm low}$, and eq.~(\ref{eqCRflux}) is arranged to guarantee that the flux is continuous across this break.  Finally, an effective low energy cutoff, $E_{\rm cut}$, is chosen, below which the flux is zero.  Despite the fact that the flux will change as cosmic rays travel into a cloud, we assume a steady state such that the spectrum is the same everywhere.  We also neglect the
possible effects of self-confinement proposed by \citet{pad05}, in which magnetohydrodynamics can spatially confine cosmic rays to given regions due to changes in the ambient density.

\begin{figure}
\epsscale{1.25}
\plotone{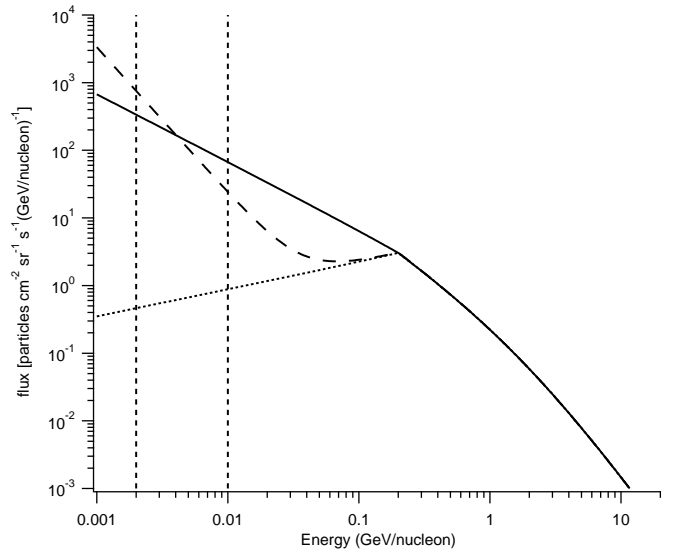}
\caption{The various cosmic-ray spectra used in this paper: dotted curve is the leaky box propagated spectrum ($\gamma_{\rm low}=0.8$); solid curve is the broken power law spectrum with $\gamma_{\rm low}=-2.0$; dashed curve is the carrot spectrum with $\alpha=-4.3$, $f=0.01$; above 0.2 GeV all three spectra are identical;
vertical dashed lines represent 2~MeV and 10~MeV low energy cutoffs.}
\label{fig2}
\end{figure}

In the case of the propagated spectrum, the momentum index
$\gamma_{\rm low}= 3 + \gamma_{\rm source} = 0.8$,
which corresponds to a source spectrum with $q(p) \propto  p^{-2.2}$ and propagation dominated by energy losses
(the ``thick-target'' approximation).
This spectrum is shown as the dotted curve in
both Figures \ref{fig1} \& \ref{fig2}.  The two vertical dashed lines in Figure \ref{fig2} represent low
energy cutoffs at 2~MeV and 10~MeV.  These were chosen
because cosmic rays with these energies have ranges roughly corresponding to the column densities of diffuse and dense clouds, respectively \citep{cra78}.
Following this reasoning, cosmic rays with $E<2$~MeV should not penetrate diffuse clouds, and so will not contribute to the ionization rate there.  Likewise, cosmic rays with energies below 10 MeV will not affect the ionization rate in dense clouds.

Another spectrum we consider is modeled after \citet{men71} and
\citet{men75a,men75b} who added a second sharply-peaked component -- dubbed a ``carrot'' -- to the propagated spectrum to give a high flux at energies of a few MeV.  The physical reasoning behind such a component is that
in addition to the propagated spectrum, there is some local source of
cosmic rays formed in weak shocks, represented by a steeper power law.
This component is given by
\begin{equation}
\phi_p(E)=f\phi_p(E_1)
\left(\frac{p(E_2)}{p(E_1)}\right)^{\gamma_{\rm high}}\left(\frac{p(E)}{p(E_2)}\right)^{\alpha},
\label{eqcarrot}
\end{equation}
where $f$ sets the flux of the carrot component to be some fraction of the
propagated spectrum at $E_2$,
and the total cosmic-ray spectrum is taken to be the sum of the propagated and
carrot components.  To ensure that the carrot does not conflict with
observations at high energies, $f$ should be relatively small ($\lesssim0.1$)
and $\alpha$ must be less than $\gamma_{\rm high}$.

In addition to the cosmic-ray spectra proposed above, we also consider several which have been used in the past for similar calculations (see \S1 and Figure \ref{fig1}). Determining the ionization rate produced by these spectra allowed us to check that our numerical integration code was working properly, and to see exactly what energy range of cosmic rays is responsible for the high ionization rate inferred from \hhh.

\section{IONIZATION ENERGETICS: THEORETICAL ESTIMATES}

Given a cosmic-ray spectrum and relevant ionization cross section, the cosmic-ray ionization rate is readily calculable.
Namely, the ionization rate of species X due to cosmic rays is given by:
\begin{equation}
\zeta_{\rm x}=4 \pi \xi_{\rm x}(1+G_{\ref{eqRateInt}})\int_{E_{\rm low}}^{E_{\rm high}}
\phi_p(E)\sigma_{\rm ion}(E) \ dE,
\label{eqRateInt}
\end{equation}
where $\phi_p(E)$ is the flux of cosmic-ray protons as a function of kinetic
energy, $\sigma_{\rm ion}(E)$ is the ionization cross section of atomic
hydrogen as a function of kinetic
energy, $G_{\ref{eqRateInt}}$ is a coefficient accounting for ionization by heavier cosmic ray
nuclei, and $\xi_{\rm x}$ converts between the primary ionization rate of atomic hydrogen $(\zeta_p)$, computed by the integral, and the total
ionization rate for a given species X.  This conversion factor includes ionization due to secondary electrons produced in the initial event, and accounts for the difference in the ionization cross section between H and X
($\zeta_{\rm x}=\xi_{\rm x}\zeta_p$).
The coefficients for atomic ($\xi_{\rm H}$)
and molecular ($\xi_2$) hydrogen are 1.5 and 2.3, respectively \citep{gla74}.
For the ionization rate calculation $G_{\ref{eqRateInt}}=0.5$ (The coefficient
$G_n$ changes based on context, and is labeled with a subscript indicating which
equation it applies to, e.g. $n=10$ in this case.  See the Appendix for a more detailed discussion).  The \citet{bet33} cross section for primary
ionization of atomic hydrogen
\begin{eqnarray}
\!\!\!\!\!\!\!\!\!\!\!\!\!\!\!\!\!\!\!\sigma_{\rm ion}
  &\!= & 2\pi (0.285) \frac{e^4}{m_e c^2 \ry } \frac{Z^2}{\beta^2}
\!\!  \left[ \ln \frac{2 m_e c^2 \beta^2}{0.048 (1-\beta^2)  \ry}
   -\! \beta^2 \!\right] \\
  & \!\!\!\!\!\!\!\!\!\!\!\! = & \!\!\!\!\!\!1.23\!\times\! 10^{-20} \frac{Z^2}{\beta^2}\!\!\left(\!\!6.2\!+\!{\rm
log}_{10}\frac{\beta^2}{1-\beta^2}\!-\!0.43\beta^2\!\!\right) \! {\rm cm^2},
\label{eqHxsec}
\end{eqnarray}
is used,
where $\beta = v/c$ is the velocity of the cosmic ray in units of the speed of
light, $\ry = 13.6$ eV is the hydrogen binding energy,
and $Z$ is the cosmic-ray charge.
Because the ionization cross section is well determined, variation of the cosmic-ray spectrum must be used to match the ionization rate inferred from
observations.

Performing a numerical integration\footnote{Integration was performed using the
{\tt qromb}, {\tt trapzd}, and {\tt polint} routines of  Numerical Recipes in FORTRAN \citep{pre92}.}
of eq.~(\ref{eqRateInt}) using
the cross section from eq.~(\ref{eqHxsec}) and the propagated spectrum ($\gamma_{\rm low}=0.8)$ form of eq.~(\ref{eqCRflux})
with $E_{\rm cut}=2$ MeV produces a cosmic-ray ionization rate of
$\zeta_2=1.4\times10^{-17}$~s$^{-1}$, about 30 times
smaller than the value inferred from \hhh.  This large discrepancy shows that a simple cosmic-ray spectrum based on the propagation of the source spectrum resulting from strong shocks is inconsistent with the ionization rate inferred from \hhh.  To reproduce the observational results then, we maintained the well-constrained high energy behavior of the cosmic-ray spectrum, but varied
$\gamma_{\rm low}$ despite the fact that this removes the initial physical motivation for the low-energy portion of the spectrum.  After several trials, we found that with $E_{\rm cut}=2$ MeV and $\gamma_{\rm low}=-2.0$ (shown as the solid curve in Figure \ref{fig2}), the above calculation gives an ionization rate of $\zeta_2=3.6\times10^{-16}$~s$^{-1}$.  However, when $E_{\rm cut}=10$ MeV is used to account for dense clouds, the calculated ionization rate is $\zeta_2=8.6\times10^{-17}$~s$^{-1}$, a bit larger than inferred values \citep{wil98,mcc99,vdvd00}.

\begin{deluxetable}{lcc}
\tablecaption{Cosmic-Ray Ionization Rates ($10^{-17}$ s$^{-1}$) \label{tblion}}
\tablehead{
 & $\zeta_2$ & $\zeta_2$ \\
 & $E_{\rm cut}=2$ MeV & $E_{\rm cut}=10$ MeV \\
Spectrum & (diffuse) & (dense)
}

\startdata

Propagated\tablenotemark{a}         &  1.4 & 1.3 \\
Broken Power Law\tablenotemark{a}   &   36 & 8.6 \\
Carrot\tablenotemark{a}             &   37 & 2.6 \\
\citet{hay61}                       &  165 & 96  \\
\citet{spi68}                       &  0.7 & 0.7 \\
\citet{nat94}                       &  260 & 34  \\
\citet{kne03}                       &  1.3 & 1.0 \\
\citet{ip85}\tablenotemark{b}       &  3.6 & 2.7 \\
\citet{her06}                       &  0.9 & 0.9 \\
\\
\hline \\
Observational Inferences & $\sim40$\tablenotemark{c} & $\sim3$\tablenotemark{d}

\enddata
\tablecomments{Ionization rates calculated are for molecular hydrogen due to a
spectrum of cosmic-ray protons and heavier nuclei with abundances greater than $10^{-5}$ with respect to hydrogen.  Factors such as the 5/3 and 1.89 used by
\citet{spi68} have been removed to calculate the primary ionization rate due to
protons, which is then multiplied by 1.5 to account for the heavy nuclei
$(1+G_{\ref{eqRateInt}})$, and 2.3 \citep{gla74} to find the H$_2$ ionization rate.}
\tablenotetext{a}{this study}
\tablenotetext{b}{via fitting function of \citet{val02}}
\tablenotetext{c}{\citet{ind07}: \hhh}
\tablenotetext{d}{\citet{vdvd00}: H$^{13}$CO$^+$}

\end{deluxetable}

We next attempted to reproduce the inferred ionization rate by using several cosmic-ray spectra in the literature.  These include spectra previously used to calculate light element abundances \citep{val02,kne03}, desorption from interstellar ices \citep{her06}, and the ionization rate\footnote{In these cases we use the same coefficients and low-energy cutoffs as for our proposed spectra, so our results differ slightly from those of the original respective papers.} \citep{hay61,spi68,nat94}.
The results of our calculations using these spectra are shown in Table \ref{tblion}, along with the results from the 3 spectra proposed in this paper.  It is interesting that none of the previous spectra are capable of reproducing the ionization rate in diffuse clouds to within even the correct order of magnitude, thus highlighting the need for the present study.

To match the ionization rate in both diffuse and dense clouds, we then turned to the carrot spectrum, which, as mentioned above, must rise faster than $\phi_p\propto p^{-2.7}$ to low energies.  Choosing $f=0.01$ and $\alpha=-4.3$ (these values optimize the ionization and light element results as discussed in \S6.1), and using $E_{\rm cut}=2$ MeV generates an ionization rate of $3.7\times 10^{-16}$~s$^{-1}$.  The carrot spectrum with these parameters is
shown in Figure \ref{fig2} as the dashed curve.  Changing the low energy
cutoff to 10 MeV to simulate a dense cloud environment predicts
$\zeta_2=2.6\times 10^{-17}$~s$^{-1}$, also in accord with observations.  This demonstrates that the steeper slope of the carrot component is better able to reproduce the roughly 1 order of magnitude difference in the ionization rate between diffuse and dense clouds.  It is also interesting to note that the large majority of ionizing cosmic rays have kinetic energies between 2 and 10 MeV: $\sim95\%$ in the case of the carrot spectrum and $\sim80\%$ for the broken power law.

\section{OTHER OBSERVABLE SIGNATURES OF LOW-ENERGY COSMIC-RAY INTERACTIONS}

As stated in \S1, cosmic rays produce light elements and gamma rays via
spallation and the excitation of nuclear states, respectively.  Like the
ionization rate, the production rates of these processes can be computed using
eq.~(\ref{eqRateInt}).  In these cases, $\sigma_{\rm ion}$ is replaced with the
relevant cross section for each process, $\xi_{\rm x} = 1$, and
$G_{\ref{eqRateInt}} = 0$.

\subsection{Light Elements}

To calculate the total production rates of the light element species $^6$Li,
$^7$Li, $^9$Be, $^{10}$B, and $^{11}$B (often collectively referred to as
LiBeB), 32 reactions were considered.  These include
the spallation (fragmentation) reactions
$[p,\alpha]+[{\rm C, N, O}] \rightarrow [{\rm LiBeB}] + \cdots$,
which make all of the LiBeB nuclides,
as well as the fusion reaction
$\alpha+\alpha \rightarrow \iso{Li}{6,7} + \ldots$,
which can only make the lithium isotopes.
Tabulated cross sections were taken
from \citet{rea84}, and for energies above $\sim 50$~MeV~nucleon$^{-1}$ the
$\alpha+\alpha$ processes were supplemented with data from \citet{mer01}.
In the case of all $\alpha$ particle processes, the fluxes of the cosmic-ray
spectra were reduced to 9.7\% of the fluxes used in the ionization calculations because of the relative solar abundance between helium and hydrogen \citep{and89}.
Using these cross sections and the spectra from \S 5, we calculated the
present-day instantaneous production rate of each species from each process.

Of course, the observed light element abundances are the result of cosmic-ray interactions with the ISM throughout the history of the Galaxy, meaning that they are dependent on the cosmic-ray history and chemical evolution of the Galaxy.  A full calculation of these effects and a comparison with LiBeB abundance evolution as traced by Galactic stars is a worthy subject of future work, but is beyond the scope of this paper.  To estimate the accumulated LiBeB abundances, we follow the original approach of \citet{rev70} to roughly quantify
the {\em solar} LiBeB abundances expected from spallation processes with our trial spectra.  Our estimate
assumes that both the cosmic-ray spectrum and CNO abundances have remained constant throughout the history of the Galaxy.  Also, we assume that once created, the light element isotopes are not destroyed.  As we know that light elements are destroyed (``astrated'') in stellar interiors, this assumption leads to further uncertainty in our model.  In addition to cosmic-ray spallation, $^7$Li and $^{11}$B are produced by other mechanisms: neutrino spallation processes in Type II SNe for both isotopes \citep{dea89,woo90},
and also primordial nucleosynthesis in the case of $^7$Li.  Both of these mechanisms will contribute to the observed abundances, but we have chosen to omit their effects with the understanding that our \li7 and \bor11 abundances should be lower than the net Galactic levels.  These effects both add to and subtract from our estimate based on a constant production rate.  Based on more detailed models which include these effects \citep{fie99,fie00,pro06} we expect our calculations of the {\it absolute} abundances to be accurate only to within factors of 2--3.  Our results appear in Table \ref{tbllibeb},
along with solar-system light element abundances as measured from meteorites and the solar photosphere.

\begin{deluxetable*}{c|c|cccc}
\tablecaption{Light Element Abundances \label{tbllibeb}}
\tablehead{
Ratio & Solar System\tablenotemark{a} & Carrot & Broken Power Law & Propagated
}

\startdata

$10^{10} \times {}^6$Li/H   & 1.5  & 2.5  & 8.2  & 1.3 \\
$10^{10} \times {}^7$Li/H   & 19   & 5.8  & 18   & 1.9 \\
$10^{10} \times {}^9$Be/H   & 0.26 & 0.35 & 0.59 & 0.33 \\
$10^{10} \times {}^{10}$B/H & 1.5  & 1.4  & 2.5  & 1.3 \\
$10^{10} \times {}^{11}$B/H & 6.1  & 3.2  & 6.4  & 2.8 \\
\li6/\be9                   & 5.8  & 7.1  & 13.9 & 4.0 \\
\bor10/\be9                 & 5.8  & 4.0  & 4.3  & 3.9 \\

\enddata
\tablecomments{For all three spectra calculations were done using $E_{\rm cut}=2$~MeV.  Calculated values were found by integrating the instantaneous rates over 10~Gyr.}
\tablenotetext{a}{abundances from \citet{and89}}
\end{deluxetable*}

As seen in Table \ref{tbllibeb},
the conventional propagated spectrum reproduces each of the
$^6$Li, $^9$Be, and $^{10}$B abundances and their ratios
well, to within 10-30\%, while severely underproducing $^7$Li and $^{11}$B.
This well-known pattern is characteristic of cosmic-ray nucleosynthesis
predictions \citep[e.g.,][and references therein]{fie01,van00}
and follows our expectations given the omission of non-cosmic-ray \li7 and \bor11 processes mentioned above.
However, as we have shown, this spectrum leads to an ionization
rate which falls far short of that required by \hhh\ data.

Turning to the spectra with low-energy enhancements and associated high ionization rates, the LiBeB production presents a mixed picture.
Here, we focus on the species with only cosmic-ray sources:
\li6, \be9, and \bor10.  In Table \ref{tbllibeb} we see that the carrot spectrum reproduces \bor10 quite well, and overproduces \li6 and \be9 by factors of 1.7 and 1.3, respectively, still well within the uncertainties of our rough calculation.  For the broken power law spectrum, however, all of these absolute abundances exceed observations by factors of about 2-5.

It is quite striking that there is only a factor of a few difference between the LiBeB abundances predicted by the propagated and carrot/broken power law spectra compared to the factor of about 30 difference for the ionization rate.  This is due to two properties of the LiBeB production cross sections. First, most of the cross sections have low-energy thresholds in the tens of MeV, meaning that the high flux in the few MeV range has no effect on LiBeB production.
Second, the cross sections do not fall off steeply as energy increases, so the cosmic-ray flux in the hundreds of MeV range (where all three spectra are identical) is important.  This contrasts with the case of ionization where cosmic rays with the lowest energies above the cutoff dominate.

However, much more significant than the absolute abundances are the {\em ratios} of these isotopes, which effectively remove the systematic uncertainties in the absolute abundances introduced by the simplicity of our model
and directly reflect the shape of the cosmic-ray spectrum.  While both of the low-energy-enhanced spectra underestimate the \bor10/\be9 ratio by about a factor of 1.5, the carrot spectrum does a much better job of reproducing the \li6/\be9 ratio; it overestimates the ratio by a factor of only $\sim1.2$ compared to the broken power law's 2.4.

\begin{figure}
\epsscale{1.25}
\plotone{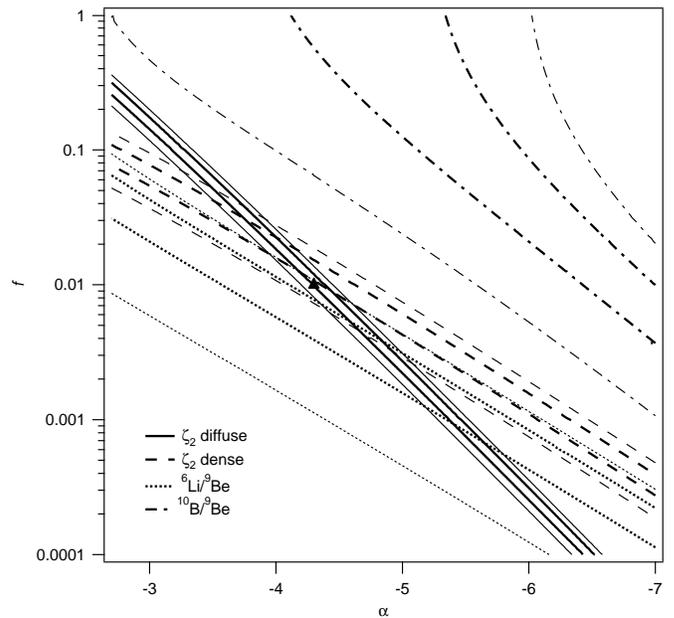}
\caption{Plot in $(f,\alpha)$ space showing how well various combinations reproduce the observed ionization rates and light element ratios.  Thicker lines represent 10\% deviation from observed values, and thinner lines show 25\% deviation.  Different styles have the following meanings: solid - $\zeta_2$ in diffuse clouds; dashed - $\zeta_2$ in dense clouds; dotted - \li6/\be9; dash-dot - \bor10/\be9.  The triangle shows the parameters chosen for the carrot spectrum: $f=0.01$, $\alpha=-4.3$
}
\label{figfalpha}
\end{figure}

This success of the carrot spectrum is not surprising though, as we chose the input parameters to best reproduce the observed ionization rates and light element ratios.  These ``optimal'' parameters were found by using various combinations of $f$ and $\alpha$ to compute the ionization rate with a 2 and 10 MeV cutoff, and the \li6/\be9 and \bor10/\be9 ratios.  Figure \ref{figfalpha} is a plot in $(f,\alpha)$ space where the contours represent deviations of 10\% and 25\% from inferred values of $\zeta_2$ in diffuse and dense clouds, and from measured meteoritic LiBeB ratios.  It is clear from Figure \ref{figfalpha} that there is an overlapping region around $f\sim0.01$ and $\alpha\sim-4.5$ where $\zeta_2$ for diffuse and dense clouds and \li6/\be9 are all within 25\% of observed values.  In making the diffuse cloud ionization rate as close to $4\times10^{-16}$~s$^{-1}$ as possible, we chose $f=0.01$ and $\alpha=-4.3$ (indicated by the triangle in Figure \ref{figfalpha}) as the parameters for the carrot spectrum.

While $\zeta_2$ and \li6/\be9 can be matched well, no combination of $f$ and $\alpha$ is able to successfully reproduce the \bor10/\be9 ratio to within 25\% simultaneously with any of the other observables.  Almost the entire range of Figure \ref{figfalpha} is within the 50\% contour of \bor10/\be9 though, so our carrot spectrum is not completely out of the question.  Indeed,
the carrot spectrum predicts almost the same \bor10/\be9 ratio as
does the propagated spectrum, so the introduction of the carrot
leaves the agreement with solar system data no worse than in the
standard case.

\subsection{Gamma Rays}

In computing the production rates of gamma-rays, 6 total reactions were used.  These include
$[p, \alpha]+\iso{C}{12} \rightarrow \iso{C}{12}^* \rightarrow \iso{C}{12}+\gamma_{4.44}$;
$[p, \alpha]+\iso{O}{16} \rightarrow \iso{O}{16}^* \rightarrow \iso{O}{16}+\gamma_{6.13}$; and $[p, \alpha]+\iso{O}{16} \rightarrow \iso{C}{12}^*+\alpha \rightarrow \iso{C}{12}+\alpha+\gamma_{4.44}$, where the
de-excitations of $\iso{C}{12}^*$ and $\iso{O}{16}^*$
produce 4.44~MeV and 6.13~MeV gamma rays,
respectively.  Cross sections for all of the above processes come from
\citet{ram79} (and references therein).
Along a line of sight with hydrogen column density $N_{\rm H}$,
the gamma-ray line intensity is
\beq
I_\gamma = N_{\rm H} \int \phi(E) \sigma_{\rm \gamma}(E) \ dE,
\label{eq:gammaflux}
\eeq
with units [photons cm$^{-2}$ s$^{-1}$ sr$^{-1}$], and
where for each line the product $\phi \sigma$ represents a sum
over appropriately weighted projectiles and targets.
If all diffuse clouds experience the same cosmic-ray flux and the column of such clouds through the Galactic plane is about
$N_{\rm H} = 10^{23}$~cm$^{-2}$,
our calculations predict that the Galactic plane should have the diffuse $\gamma$-ray fluxes shown in Table \ref{tblgamma}.
Assuming the intensity in eq.~(\ref{eq:gammaflux})
is uniform in the Galactic plane, we integrated over the solid angle within $|l|<30^{\circ}$, $|b|<10^{\circ}$ to find the total flux in the central radian of the Galaxy (a typical region over which diffuse $\gamma$-ray fluxes are quoted, with units [cm$^{-2}$ s$^{-1}$ rad$^{-1}$]).  Also shown in Table \ref{tblgamma} is the large-scale sensitivity of the {\it INTEGRAL} spectrometer at $\sim5$~MeV \citep{tee06}.  Our predicted fluxes are below currently available detector limits of {\it INTEGRAL}.  Thus the presence of low-energy cosmic-rays sufficient to give the ionization levels required by \hhh\ does not violate gamma-ray constraints.

\begin{deluxetable*}{l|c|ccc}
\tablecaption{Diffuse Gamma-Ray Flux from the Central Radian\tablenotemark{a} ($10^{-5}$ s$^{-1}$ cm$^{-2}$ rad$^{-1}$)
\label{tblgamma}}
\tablehead{
Energy & {\it INTEGRAL} sensitivity & Carrot & Broken Power Law & Propagated
}

\startdata
4.44 MeV & 10 & 3.0 & 8.3 & 0.9 \\
6.13 MeV & 10 & 2.4 & 5.9 & 0.4 \\
\enddata
\tablecomments{Predicted fluxes for the 4.44 and 6.13 MeV $\gamma$-ray lines using our carrot, broken power law, and propagated spectra.  All calculations were done using $E_{\rm cut}=2$~MeV.  Also shown are the most directly comparable sensitivities of the {\it INTEGRAL} spectrometer given by \citet{tee06}}
\tablenotetext{a}{For the central radian we considered uniform emission within $|l|<30^{\circ}$ and $|b|<10^{\circ}$}
\end{deluxetable*}

Indeed, the gamma-ray line predictions in Table  \ref{tblgamma} lie tantalizingly close to present limits.  While this does not provide a test of the predicted cosmic-ray spectra at present, it may be possible that {\it INTEGRAL} itself, and certainly the next generation gamma-ray observatory, will have the ability to detect these lines.  In any case, the results
show that our proposed spectra are not inconsistent with observations.

\subsection{Energetics}

Similar to the calculations in \S3, we can determine the energy budget for a given cosmic-ray spectrum.  Unlike the previous calculation though, in this case the shape of the spectrum is important as we compute the energy loss rate of all of the cosmic rays in the spectrum.  This is done via the usual
Bethe-Bloch expression for energy loss $dE$ per unit
mass column $dR = \rho dx = \rho v dt$:
\begin{equation}
\frac{dE}{dR}  =\frac{4\pi zZ^2e^4}{A\langle m\rangle m_ev^2}
\left[\ln\left(\frac{2\gamma^2m_ev^2}{I}\right)-\beta^2\right],
\label{eqeloss}
\end{equation}
which is closely related to the ionization cross section above
(eq.~\ref{eqHxsec}).
Here we use $z=Z=A=1$, $\langle m\rangle =1.4m_{\rm p}$, and
$I=13.6$~eV (see \citet{pro03} for a complete description of the variables
involved).
The rate of cosmic-ray energy loss per unit mass of neutral hydrogen
is
\beq
\label{eq:dLdM}
\frac{L_{\rm CR}}{M_{\rm H}}
 = 4\pi (1+G_{\ref{eq:dLdM}}) \int_{E_{\rm low}}^{E_{\rm high}} \phi_p(E) \ \frac{dE}{dR} \ dE,
\eeq
where $G_{\ref{eq:dLdM}}=0.1$ (see the Appendix).
Again using a Galactic hydrogen mass of
$M_{\rm H} = 5\times 10^{9} \msol$,
the carrot and broken power law spectra require energy inputs of
$L_{\rm CR} = 0.18\times 10^{51}$ erg (100~yr)$^{-1}$ and
$0.17\times 10^{51}$ erg (100~yr)$^{-1}$, respectively.  These represent large fractions of the total mechanical energy released in SNe.  Like in \S3, they are also consistent with the $0.5\times 10^{51}$ erg (100 yr)$^{-1}$ found by \citet{fie01} which accounted for both energy
needed for ionization of the ISM and escape from the Galaxy.  Finally, we note that the large amounts of energy and high acceleration efficiencies required may help to resolve the superbubble ``energy crisis'' described by \citet{but08}.

\begin{figure}
\epsscale{1.25}
\plotone{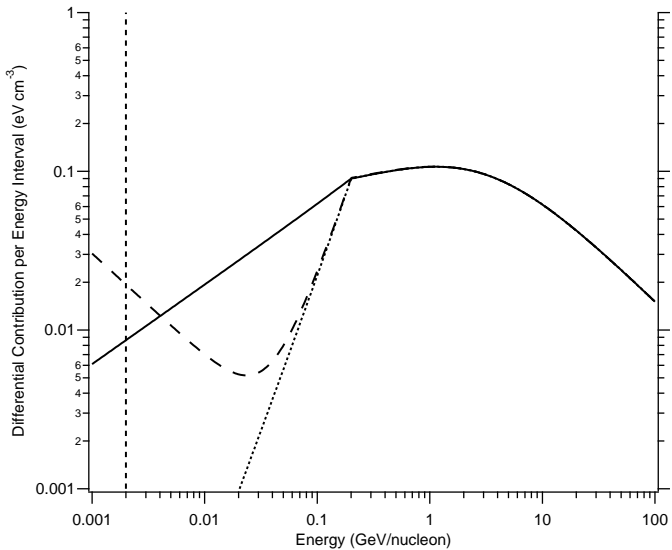}
\caption{Contribution to the energy density of cosmic rays as a function of
kinetic energy per nucleon.  As in Figure \ref{fig2}, the dotted curve is the
leaky box propagated spectrum, the solid curve is the broken power law spectrum,
the dashed curve is the carrot spectrum, and the vertical dashed line shows the
2~MeV low energy cutoff.  The vertical axis is given by
$Ed\epsilon/dE$, where $d\epsilon/dE=4\pi E\phi_p(E)/v(E)$.}
\label{figedense}
\end{figure}

Beyond the total input energy requirement, each cosmic-ray spectrum will also have a particular energy density and pressure.  Energy density can be calculated from
\begin{equation}
\label{crEdense}
\epsilon_{\rm CR}
 =  4\pi (1+G_{\ref{crEdense}}) \int_{E_{\rm low}}^{E_{\rm high}} E \frac{\phi_p(E)}{v(E)} dE,
\end{equation}
where $v(E)$ is the velocity and $G_{\ref{crEdense}}=0.42$ (see Appendix), and pressure from
\begin{equation}
\label{crPress}
P_{\rm CR}
 =  \frac{4\pi}{3} (1+G_{\ref{crPress}}) \int_{E_{\rm low}}^{E_{\rm high}} \phi_p(E) p(E) dE,
\end{equation}
where $p(E)$ is again relativistic momentum, and $G_{\ref{crPress}}=0.42$.  Performing these
calculations with $E_{\rm cut}=2$~MeV results in energy densities of
0.77~eV~cm$^{-3}$ and 0.89~eV~cm$^{-3}$, and pressures $(P_{\rm CR}/k_B)$ of
4300~K~cm$^{-3}$ and 5200~K~cm$^{-3}$ for the carrot and broken power law
spectra, respectively. Both pressures are in rough accord with the average
thermal pressure in the diffuse ISM of $(P/k_B)=2700$~K~cm$^{-3}$ reported by
\citet{jen07}.  The energy densities in both spectra, however, are about one
half of the value reported by \citet{web98}.  While this result may at first
seem counterintuitive, it is best clarified graphically by Figure \ref{figedense}.  Here, it is shown that cosmic rays with energies between about 0.1~GeV and 10 GeV completely dominate in contributing to the energy density.  This was previously demonstrated by the analogous plot (Fig. 7) in \citet{web98}, from which the author concluded that low energy components, such as those proposed in this paper, would have little effect on the cosmic-ray energy density.  As for why our spectra have lower energy densities, this is almost entirely dependent on the flux normalization at higher energies.  At about 1~GeV the fluxes in our spectra are about one half that of the local interstellar spectrum used by \citet{web98}, thus resulting in the corresponding factor of 2 difference in energy densities.

\subsection{Cloud Heating}

One further effect that cosmic rays have is to heat the ISM via energy lost
during the ionization process.  On average, 30 eV are lost by a cosmic ray
during each ionization event \citep{cra78}.  Using $\zeta_2=4\times 10^{-16}$ s$^{-1}$ and the corresponding $\zeta_{\rm H}=1.5\zeta_2/2.3$ and assuming that the number density of atomic hydrogen is roughly equal to that of molecular hydrogen, and that all of the lost energy eventually goes into heating, we find a heating rate due to cosmic rays of
$\Gamma_{\rm CR}=3\times 10^{-26}$ erg s$^{-1}$ (H atom)$^{-1}$.  This can be
compared to the heating rate due to the photoelectric effect calculated by
\citet{bak94} of $\Gamma_{\rm PE}=1.5\times 10^{-25}$ erg s$^{-1}$ (H atom)$^{-1}$
for the diffuse cloud sight line toward $\zeta$ Oph.  The heating rate caused
by cosmic-ray ionization is about 5 times smaller than the heating rate due to the
photoelectric effect, demonstrating that even with such a high ionization rate, cosmic rays do
not significantly alter the heating rate in diffuse clouds.  This large difference in heating
rates is further illustrated by Figure 10 of \citet{wol03}.  Because the high flux of low
energy cosmic rays we use will not dominate cloud heating, our spectra do not imply cloud temperatures that are inconsistent with observations.

\section{DISCUSSION}

\subsection{Cosmic-Ray Spectra}

As described in \S4, the spectrum with seemingly the best physical motivation is one that arises from the propagation of particles accelerated by strong shocks in supernova remnants.  This spectrum follows a $p^{-2.7}$ relationship above a few hundred MeV, matching observations, and a $p^{0.8}$ relationship below a few hundred MeV in the ionization-dominated regime.  A similar behavior is apparent in the spectra used by \citet{hay61}, \citet{spi68}, and \citet{her06}
(see Figure \ref{fig1}), except they follow power laws closer to $p^2$ at low energies.  Even the more sophisticated models such as those considering re-acceleration \citep{shi06} or distributed cosmic-ray sources and a Galactic wind \citep{ler82} follow these general power laws.  Because they decrease at low energy though, all of these spectra (except \citet{hay61} which does not begin decreasing until $E\leq10$~MeV) are unable to provide enough flux at the energies where ionization is most efficient, and thus cannot generate the high ionization rate inferred from \hhh.  It seems then that the propagation of cosmic rays accelerated by SNR's (with test-particle power-law spectra) cannot explain the high flux of low energy cosmic rays necessary in our spectra.
However, plentiful evidence supports the idea that high-energy ($\ga 1$ GeV) Galactic cosmic rays {\em do} originate in supernovae: synchrotron emission in supernova remnants indicates electron acceleration and strongly suggests ion acceleration, and Galaxy-wide cosmic-ray energetics are in line with supernova expectations and difficult to satisfy otherwise.  For this reason we retained the propagated cosmic-ray spectrum to represent the supernova-accelerated Galactic cosmic-ray component at high energies ($\ga 1$ GeV), but also considered {\em additional} cosmic-ray components which dominate at low energies where the ionization efficiency is high.

While our carrot and broken power law spectra do not follow the
conventional strong-shock thick-target
$\phi \propto p^{0.8}$ relationship in the low-energy
regime, recent studies are beginning to find possible sources for the proposed high flux of low energy cosmic rays.  Weak shocks will accelerate cosmic rays with preferentially steeper power laws (see \S4), and may be ubiquitous in the ISM, caused by star forming regions, OB associations, and even low mass stars like the sun.  \citet{sto05} investigated the flux of termination shock protons at the heliosheath using the {\it Voyager} spacecraft.  Their study showed that at low energies, $0.5~{\rm MeV}\lesssim E \lesssim 3.5~{\rm MeV}$, $\phi_p \propto E^{-1.4}$ $(\propto p^{-2.8})$, and that from about 3.5-10 MeV, this relationship steepened to $\phi_p \propto E^{-2}$ $(\propto p^{-4})$.
These so-called ``anomalous'' cosmic rays (ACRs) clearly demonstrate that real sources exist
in nature which can produce a high particle flux at low energies, though
these measurements and the anomalous cosmic rays themselves are located in the
heliosphere, not interstellar space.
\citet{sch08} modeled the effects from the astropauses of all F, G, and K type stars in the Galaxy to find a power of $2.2\times 10^{49}$ erg (100 yr)$^{-1}$ in ACRs.  This accounts for only about 10\% of the power needed to produce the ionization rate inferred from \hhh.
However, their analysis did not include the effects of winds from the much more luminous O and B stars.  To our knowledge, no study has computed the interstellar cosmic-ray spectrum arising from the ACRs of all stellar winds in the Galaxy, but it may indeed be an important contribution to the flux of low energy cosmic rays.

Another intriguing possibility is that supernova shocks are considerably more efficient at low-energy particle acceleration than simple test-particle results would indicate. Indeed, theoretical non-linear shock acceleration calculations
\citep[e.g.,][]{kan95,ber99,bla02} do predict that low-energy particles
have higher fluxes than in the test-particle limit. This result goes in the
right direction qualitatively. However, published source spectra we are aware of do not appear steep enough at low energies -- particularly after ionization losses are taken into account in propagation -- to reproduce the observed ionization rate. It remains an interesting question whether more detailed nonlinear calculations, focussing on the low-energy regime, might provide a solution to the ionization problem; in this case, ionization rates and gamma-ray lines would become new probes of feedback processes in supernova remnants.

In addition to protons and heavy nuclei, it has been proposed that cosmic-ray electrons may make a significant contribution to the ionization rate. \citet{web98} showed that the local interstellar spectrum of cosmic-ray electrons produces a primary ionization rate of $\sim2\times10^{-17}$~s$^{-1}$ when considering energies above $\sim2$ MeV.  While this is roughly equal to the ionization rate inferred for diffuse clouds at that time, electrons only account for about 10\% of the primary ionization rate inferred from \hhh\ \citep{ind07}.  As a result, we have neglected the effect of cosmic-ray electrons in the present study.  It is worth noting, however, that low-energy cosmic-ray electrons are
probed by very low-frequency radio emission, and indeed cosmic-ray electron emission provides a major foreground for present and future facilities aimed at the measurement of cosmological 21-cm emission from high-redshift sources, including LOFAR \citep{lofar} and the Square Kilometer Array \citep{ska}.
Such observations should provide a detailed picture of low-energy cosmic-ray electrons, whose behavior can in turn be compared to the proton and nuclear components probed by the other observables considered in this paper.

Finally, we have found that the carrot spectrum
produces \li6/\be9 in good agreement with solar system data,
and a \bor10/\be9 ratio which is almost identical to the
standard propagated result but which is somewhat low with respect
to the solar system value.
To the extent that the isotope ratios are not in perfect
agreement with solar system data, one possible explanation
could be time variations of the cosmic-ray spectral shape over
Galactic history.  If supernovae are the agents of cosmic-ray
acceleration, then presumably strong shocks will always
lead to high-energy source spectra with $\gamma_{\rm source} \sim -2.2$
as we have today.  However, the low-energy component responsible
for the carrot derives from weaker shocks which in turn
may reflect time-dependent properties of, e.g., star-forming
regions.  Moreover, cosmic-ray {\em propagation}
is much more sensitive to the details of the interstellar environment,
particularly the nature of Galactic magnetic fields.
\citet{pra93} suggested that such variations in the early Galaxy
might explain the B/Be ratios in primitive (Population II) halo
stars.
Similarly, if such variations were present in the later phases
of Galactic evolution (e.g., during major merger events)
then it is possible that the propagated cosmic-ray spectra
could have differed substantially.  The resulting
LiBeB contributions could alter the ratios from the simple
time-independent estimates we have made.

Another possible explanation to bring the theoretical LiBeB ratios into better agreement with observations would arise if the LiBeB isotopes suffer significantly different amounts of destruction (astration) in stellar interiors. Because the binding energies are in the hierarchy
$B(\li6) < B(\be9) < B(\bor10)$, there should be a similar ranking of the
fraction of the initial stellar abundance of these isotopes
which survives to be re-ejected at a star's death.
If the amount of \li6 destroyed is greater than \be9, which is in turn greater than \bor10, then the results from our carrot spectrum may be correct before accounting for astration.  Assuming this preferential destruction decreases our calculated \li6/\be9 and increases \bor10/\be9, moving both closer to the measured ratios.  That said, conventional stellar models
\citep[e.g.,][]{sac99} and their implementation in Galactic chemical evolution \citep{ali02} find different, but still small, survival fractions for all isotopes, $\la 10\%$ for \bor10. As a result, this scenario would seem to require large upward revisions to the survival of \be9 and \bor10 in stars.

\subsection{Astrochemistry}

Gas phase chemistry in the ISM is driven by ion-molecule reactions.  Photons with $E>13.6$~eV are severely attenuated in diffuse and dense clouds, meaning that cosmic rays are the primary ionization mechanism in such environments.  As a result, the cosmic-ray ionization rate has a large effect on the chemical complexity of the cold neutral medium.  In fact, it has a direct impact on the abundances of \hhh, OH, HD, HCO$^+$, and H$_3$O$^+$, to name a few molecules.
This makes the cosmic-ray ionization rate an important input parameter for astrochemical models which predict the abundances of various atomic and molecular species.

However, based on the current theoretical study it seems that instead of having a uniform Galactic value, the cosmic-ray ionization rate should vary between sight lines.  This has to do with the source behind cosmic-ray acceleration.  While it has typically been assumed that low energy cosmic rays are accelerated in supernovae remnants, the spectra making this assumption were unable to reproduce the ionization rate inferred from \hhh.  Instead, a low energy carrot was required, most likely produced by particles accelerated in weaker, more localized shocks.  Unlike the SNe cosmic rays which are assumed to diffuse throughout the Galaxy, cosmic rays accelerated in weak local shocks could lead to significant enhancements in the local ionization rate.  Observations of \hhh\ have shown that the cosmic-ray ionization rate is in fact variable between different diffuse cloud sight lines.  The H$_2$ ionization rates toward $\zeta$~Per and X~Per are about $7\times10^{-16}$~s$^{-1}$ \citep{ind07}, while $3\sigma$ upper limits toward $\zeta$~Oph and $o$~Sco are as low as $1.6\times10^{-16}$~s$^{-1}$ and $1.2\times10^{-16}$~s$^{-1}$, respectively\footnote{These limits are based on observations performed after the publication of \citet{ind07}, and will be described in more detail in a future publication}.  Like the results of \citet{van06} for dense clouds, this demonstrates that the cosmic-ray ionization rate can vary significantly between diffuse clouds as well.  Also, it suggests that instead of searching for or adopting a ``canonical'' ionization rate, sight lines must be evaluated on a case-by-case basis.

In order to test the theory that low energy cosmic rays are primarily accelerated by localized shocks, we propose two complementary observational surveys.  First, the ionization rate should be inferred along several diffuse cloud sight lines which are surrounded by different environments.  Observations of \hhh\ in sight lines near OB associations and sight lines near low mass stars should provide data in regions near and far from energetic sources.  We expect the sight lines near more energetic regions to show higher ionization rates than those near less energetic regions.  If observations confirm these predictions, then we will be able to more confidently conclude that most of the low energy ionizing cosmic rays are accelerated in localized shocks.

The second set of observations we propose examines the ionization rate in regions of varying density.  Following the reasoning of \S4 where we assume that the lower energy cosmic rays do not penetrate denser clouds, the ionization rate should be inversely related to the cloud density.  Observing \hhh\ in diffuse clouds, dense clouds, and in sight lines with intermediate densities should provide us with a range of ionization rates.  We can then use the carrot spectrum with appropriate low energy cutoffs to reproduce the inferred ionization rates from each environment, thus constraining the slope of the carrot component.  This slope will then allow us to roughly determine the strength (or rather weakness) of the shock necessary to produce such a steep power law, and thus infer the source of the shock.

\section{CONCLUSIONS}

Three theoretical low energy cosmic-ray spectra have been examined, all of
which are consistent with direct cosmic-ray observations at high energies.
We first adopted the standard  $q \propto p^{-2.2}$ source spectrum
resulting from cosmic-ray acceleration by strong shocks in supernovae.
The propagated version of this spectrum produced an ionization rate about 30 times smaller than that inferred from \hhh, thus demanding that additional low-energy cosmic-rays be responsible for the observed ionization.

We thus studied the effects of {\em ad hoc} but physically motivated
low-energy cosmic-ray components.  We found that
the carrot and broken power law spectra could be fashioned so as to reproduce observed results for diffuse clouds.  Out of these two, the carrot spectrum did a much better job of matching the ionization rate in dense clouds.  Unlike the broken power law, the carrot spectrum was also capable of matching observed light element abundances to within a factor of 2 for the three isotopes produced solely by cosmic-ray spallation.
These results are well within the expected uncertainties of
LiBeB Galactic chemical evolution.

Predictions for the gamma-ray line fluxes for both the carrot and broken power law were below the limits of current
instruments, so these spectra are not inconsistent with data.
Indeed, our estimates are close to the detection limits of {\it INTEGRAL},
and thus motivate a search for the 4.44 MeV and 6.13 MeV lines
(or limits to their intensity)
in the gamma-ray sky.

Finally, the
energy necessary to accelerate all of the cosmic rays in these spectra is
about $0.2\times 10^{51}$ erg (100 yr)$^{-1}$.  This is a substantial fraction of the mechanical energy released in supernovae explosions, although based on our results it may be necessary that much of this energy come from weak shocks in order to produce a high flux of low energy cosmic rays.

Together, all of these calculations demonstrate that the proposed carrot
spectrum is consistent with astrochemical and astrophysical constraints.
Whether or not low-energy cosmic rays take precisely this spectral form,
at the very least this example serves as a proof by construction that
one can make cosmic-ray models which contain low-energy enhancements required by the high ionization rate inferred from \hhh, while not grossly violating other observational constraints.  This motivates future work which looks in more detail at the impact of low-energy cosmic rays.

The authors would like to thank T. Oka for insightful discussions.  NI and BJM have been supported by NSF grant PHY 05-55486.

\appendix

\section{Appendix:  Effects of Heavy Cosmic-Ray Nuclei}

For the sake of clarity, discussions of the cosmic-ray spectra in the body of
the paper focused only on the proton spectrum.  Our calculations, however,
included the effects of heavier nuclei cosmic rays as a coefficient
in eqs.~(\ref{eqRateInt}, \ref{eq:dLdM}, \ref{crEdense}, \& \ref{crPress}).  This appendix discusses in more detail the
calculations that went into determining the coefficients $G_n$.

We assume that all heavy cosmic-ray nuclei have the same spectral shape as
protons, but that their fluxes are shifted down by their respective relative
abundances to hydrogen (e.g. $\phi_{\rm He}(E)=0.097\phi_p(E)$).  With this
assumption, the contribution to the ionization rate due to heavy nuclei can be
calculated from
\begin{equation}
G_{\ref{eqRateInt}}=\sum_{i}Z_i^2g_i,
\label{eqa1}
\end{equation}
where $Z_i^2$ is the charge which comes from eq.~(\ref{eqHxsec}), $g_i$ is the fractional abundance with respect to hydrogen, and the index $i$ sums over all species with solar abundances $g_i>10^{-5}$ ($^{4}$He, $^{12}$C, $^{14}$N, $^{16}$O, $^{20}$Ne,
$^{24}$Mg, $^{28}$Si, $^{32}$S, $^{56}$Fe \citep{mey98}).  Performing this
summation results in the value of $G_{\ref{eqRateInt}}=0.5$ used in the ionization calculations of \S5.
In \S6.3, however, the energy loss per unit hydrogen mass
(eq.~\ref{eq:dLdM}) is controlled by the particle energy loss per
unit mass column $dE/dR \propto Z^2/A$ (eq.~\ref{eqeloss}).
This changes the heavy nuclei coefficient to
\begin{equation}
G_{\ref{eq:dLdM}}=\sum_{i}\frac{Z_i^2g_i}{A_i},
\label{eqa2}
\end{equation}
where the atomic mass, $A_i$, is now included because of eq.~(\ref{eqeloss}).
As a result, for the energy loss rate calculation $G_{\ref{eq:dLdM}}=0.1$.  Also in \S6.3, the
energy density (eq. \ref{crEdense}) and pressure (eq. \ref{crPress})
calculations both require the coefficient
\begin{equation}
G_{\ref{crEdense}}=G_{\ref{crPress}}=\sum_{i}A_i g_i.
\label{eqa3}
\end{equation}
Here, $A_i$ is required because $E$ and thus $p(E)$ are both in units of per
nucleon throughout the paper.  In this case, $G_{\ref{crEdense}}=G_{\ref{crPress}}=0.42$ for both the cosmic-ray energy density and
pressure calculations.

However, if the relative abundances of Galactic cosmic rays (GCRs) measured at
higher energies \citep{mey98} are used instead of solar abundances, the above
coefficients change.  This is because the abundances of most heavy nuclei are
enhanced in GCRs when compared to solar.  For the case of ionization, $G_{\ref{eqRateInt}}$
becomes 1.4, making heavy nuclei more important than protons. Because the
integral is multiplied by $(1+G_{\ref{eqRateInt}})$ though, the overall difference in the
ionization rate between using solar and GCR abundances is only a factor of
2.4/1.5=1.6.  Using GCR abundances to calculate the energy loss rate changes
$G_{\ref{eq:dLdM}}$ by a negligible amount, from 0.1 to 0.11.  Finally, GCR abundances only
change $G_{\ref{crEdense}}$ and $G_{\ref{crPress}}$ from 0.42 to 0.46 for the energy density and pressure calculations.
Despite the fact that heavy nuclei are measured to be more abundant in Galactic
cosmic rays than in the solar system, we have chosen to use solar abundances in
our calculations.
This is because the high energy cosmic rays observed are accelerated in
metal-rich SNRs, while the high flux of low energy cosmic rays is most likely
due to weak shocks associated with low mass stars and the ISM.  Due to this
source difference, we find it justifiable to use solar abundances instead of the
measured high energy GCR abundances.














\end{document}